\documentclass[conference]{IEEEtran}
\usepackage{setspace}
\usepackage{multirow,booktabs}
\usepackage{graphicx,epstopdf}
\usepackage[cmex10]{amsmath}
\usepackage{amssymb}
\usepackage{amsthm}

\usepackage[ruled,vlined]{algorithm2e}
\usepackage{algorithmic}

\newtheoremstyle{theroremthm}
{1pt}
{1pt}
{}
{}
{\bf}
{:}
{.5em}
{}

\theoremstyle{theroremthm}

\usepackage{xpatch}
\makeatletter
\xpatchcmd{\proof}{\@addpunct{.}}{\@addpunct{:}}{}{}
\makeatother

\usepackage{amsfonts}
\usepackage{fmtcount}
\DeclareGraphicsExtensions{ps,eps,pst,pdf}
\usepackage{cite}

\usepackage{stfloats}
\usepackage{color}

\usepackage{subcaption}

\usepackage{amsmath,bm}

\DeclareMathOperator{\sinc}{sinc}

\usepackage{soul}  
\usepackage[font=small]{caption}

\newcommand{\qa}{{\bf a}}

\newcommand{\qh}{{\bf h}}

\newcommand{\qn}{{\bf n}}

\newcommand{\qu}{{\bf u}}

\newcommand{\qy}{{\bf y}}

\newcommand{\qH}{{\bf H}}
\newcommand{\qI}{{\bf I}}

\newcommand{\qR}{{\bf R}}

\newcommand{\qzero}{{\bf 0}}

\newcommand{\qPhi}{{\boldsymbol \Phi}}

\newcommand{\qmathbbC}{{\mathbb C}}
\IEEEoverridecommandlockouts

\usepackage{atbegshi}
\AtBeginDocument{\AtBeginShipoutNext{\AtBeginShipoutDiscard}}

\begin{document}
\title{\LARGE{Model-free Optimization and Experimental Validation of RIS-assisted Wireless Communications under Rich Multipath Fading}}
\author{Tianrui Chen, Minglei You, Yangyishi Zhang, Gan Zheng, Jean Baptiste Gros, \\Geoffroy Lerosey, Youssef Nasser, Fraser Burton, and Gabriele Gradoni}
\thanks{Tianrui Chen and Minglei You are with University of Nottingham (Email: \{Tianrui.Chen, Minglei.You\}@nottingham.ac.uk), Gan Zheng is with University of Warwick (Email:  gan.zheng@warwick.ac.uk), Yangyishi Zhang and Fraser Burton are with Applied Research, BT (Email: \{yangyishi.zhang, fraser.burton\}@bt.com), Jean Baptiste Gros, Geoffroy Lerosey and Youssef Nasser are with Greenerwave (Email: \{jean-baptiste.gros, geoffroy.lerosey, youssef.nasser\}@greenerwave.com), and Gabriele Gradoni is with University of Surrey (Email: g.gradoni@surrey.ac.uk).

This work was supported in part by Digital Nottingham Project, by CELTIC-NEXT C2019/2-5: Artificial Intelligence Enabled Massive Multiple-Input Multiple-Output (AIMM) Project, by the Royal Society under Grant IEC$\backslash$NSFC$\backslash$223152, and by the European Space Agency under Grant 12893212.
Gabriele Gradoni has been supported by the EU H2020 RISE-6G Project under Grant 101017011, and by the Royal Society Industry Fellowship under Grant INF$\backslash$R2$\backslash$192066. 
}
\maketitle
\begin{abstract}
Reconfigurable intelligent surface (RIS) devices have emerged as an effective way to control the propagation channels for enhancing the end-users' performance. However, RIS optimization involves configuring the radio frequency response of a large number of radiating elements, which is challenging in real-world applications due to high computational complexity. In this paper, a model-free cross-entropy (CE) algorithm is proposed to optimize the binary RIS configuration for improving the signal-to-noise ratio (SNR) at the receiver. 
One key advantage of the proposed method is that it only requires system performance indicators, e.g., the received SNR, without the need for channel models or channel state information.
Both simulations and experiments are conducted to evaluate the performance of the proposed CE algorithm. This study provides an experimental demonstration of the channel hardening effect in a multi-antenna RIS-assisted wireless system under rich multipath fading.

\end{abstract}
\begin{IEEEkeywords}
Reconfigurable intelligent surface (RIS), Phase shifts optimization, Cross-entropy algorithm, Channel hardening.
\end{IEEEkeywords}
\IEEEpeerreviewmaketitle
\vspace{-0.5cm}
\section{Introduction}
\vspace{-0.2cm}
Reconfigurable intelligent surfaces (RISs), with the potential to implement smart radio environments, have emerged as an energy-efficient and cost-effective technology to create a desirable propagation channel between the transceivers by phase shifts of imping signals \cite{basharat2021reconfigurable}.
Compared to the conventional diagonal RIS design, the novel beyond diagonal RIS structure increases the flexibility of the RIS phase shifts and enhances the channel gain accordingly \cite{li2022reconfigurable} \cite{li2022beyond}. 
However, the phase shifts optimization is a challenging problem due to a large number of discrete variables being optimized. Many researchers have made efforts on such optimizations for RIS-aided wireless communications. 
A genetic algorithm was proposed in \cite{zhi2022power} to optimize the phase shifts of RIS in massive multiple-input-multiple-output (MIMO) system, but it requires statistical channel state information (CSI) estimation \cite{ahmed2005parameter}. 
A Monte-Carlo (MC) based random energy passive beamforming was proposed for RIS-assisted wireless
power transfer systems \cite{lu2022monte}, and this method can be implemented without CSI. Additionally, other MC sampling based algorithms, such as simulated annealing (SA) \cite{mohammed2022reconfigurable} and Metropolis–Hastings (MH) \cite{deng2021metropolis}, are also capable to solve such RIS optimization problem.
The aforementioned state-of-the-art works provided only numerical simulations but not experimental evaluations. 
Some works made efforts on experimental evaluations of the RIS optimization in wireless communications, e.g., multiple links in indoor environment \cite{lodro2022experimental} and millimeter wave beamforming \cite{gros2021reconfigurable}. 

In the Massive MIMO scenario, the essence of the channel hardening is that the channel variations reduce as more antennas are added \cite{bjornson2017massive}. The state-of-the-art works focus on the numerical studies about the channel hardening in RIS-aided communications such as in Rayleigh fading model \cite{Bjornson2021} and under large intelligent surface scenario \cite{Jung2020}. However, to the authors’ best knowledge, existing works have not provided their experimental validations of the channel hardening effect in the RIS-assisted communications \cite{gros2022multi}.
The main contributions of this work are summarized as follows:
\begin{itemize}
\item A cross-entropy (CE) algorithm is proposed to optimize the RIS configuration for SNR enhancement. It is a model-free algorithm that can be applied to arbitrary use cases and scenarios. Additionally, it only requires the received SNR as performance feedback and does not need channel model or CSI.
\item In addition to numerical simulations, experimental evaluations are performed for the proposed CE algorithm in a RIS-aided communication system operating within an electromagnetic reverberation chamber (RC) that emulates rich multipath fading \cite{primiani2020reverberation}.
\item In contrast to the channel hardening studied by numerical simulations in the literature, this work provides experimental validations of the channel hardening effect in the RIS-assisted communication system.
\end{itemize}

The rest of this paper is organized as follows. Section \ref{sec_sys} introduces the system model and problem formulation of a RIS-aided communication system. The proposed model-free CE algorithm for the RIS phase shifts optimization is presented in Section \ref{sec_ce}. Section \ref{sec_setting} introduces both the simulation and the experiment settings, followed by the corresponding results in Section \ref{sec_result}. Finally, conclusions are drawn in Section \ref{sec_con}.

\section{System Model and Problem Formulation}
\label{sec_sys}
\begin{figure}[!htb]
\centerline{\includegraphics[trim=0cm 0cm 0cm 0cm,clip,width=0.36\textwidth]{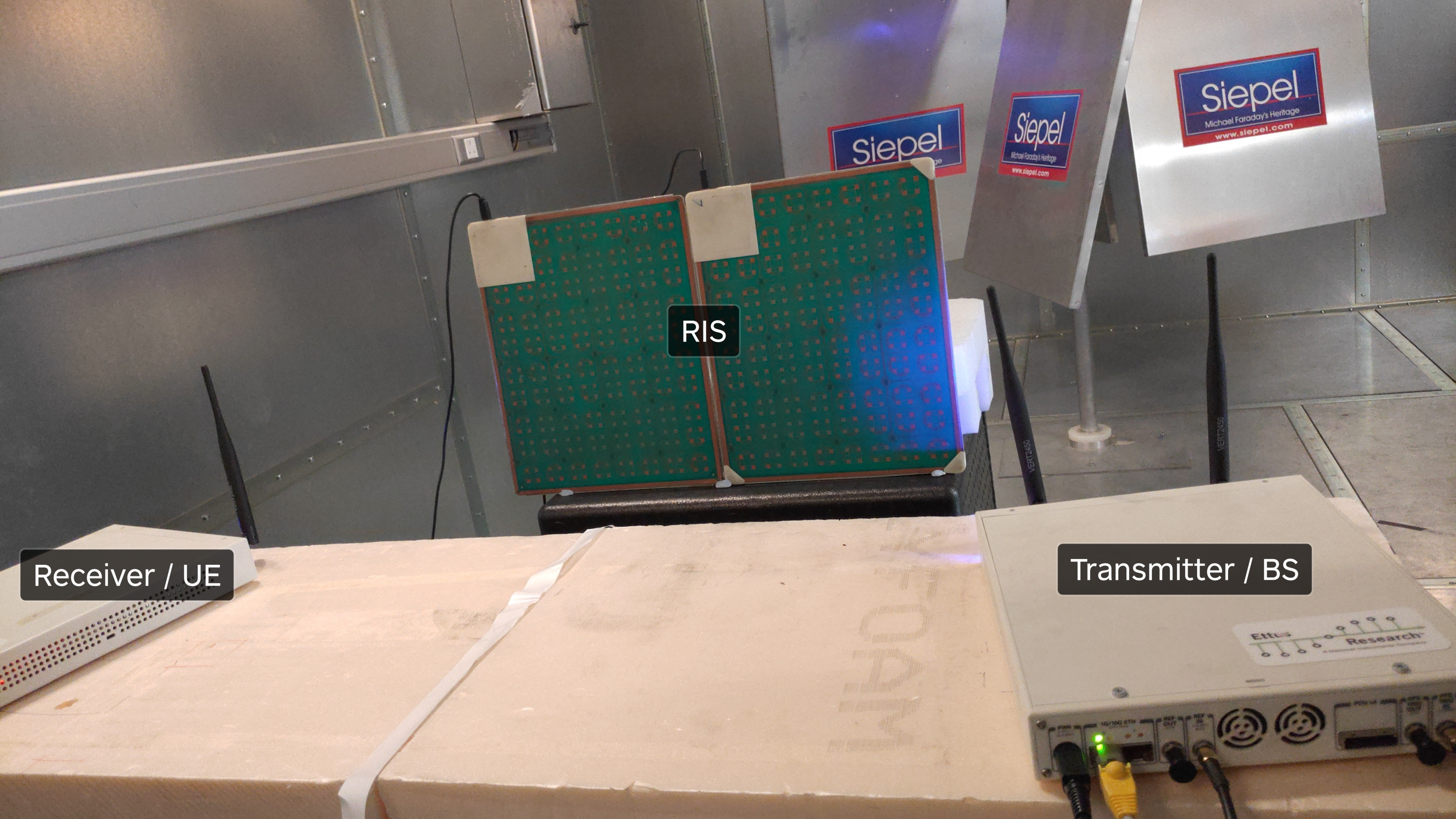}}
\caption{\footnotesize{An illustration of the RIS-assisted communication system with $N = 152$ elements. The RIS elements are dipoles which cause omnidirectional scattering, and the metallic cavity whose symmetry is broken by the stirrer supports rich multipath fading. With a strong line of sight (LOS), the fading channel gain is expected to be Ricean distribution.}}
\label{sys_model}
\end{figure}

A RIS-assisted communication system is considered as illustrated in Fig. \ref{sys_model}, where the RIS is equipped with $N$ reflective elements located between the base station (BS) with a MIMO $M$-antenna array and the single-antenna user equipment (UE).

Let $\qh_\text{bu} \in  \qmathbbC^{M \times 1}$, $\qh_\text{ru} \in  \qmathbbC^{N \times 1}$ and $\qH_\text{br}\in  \qmathbbC^{M \times N}$ denote the channels between BS-UE, RIS-UE and BS-RIS, respectively. RIS reflection coefficients are denoted as the diagonal matrix $\qPhi \in \qmathbbC^{N \times N}$, whose element $\Phi_{nn} = e^{j\phi_n}, \forall n =1, \dots, N$. 
Let $\bm{x} = \{x_n\}_{n=1}^{N}$ denote the phase shifts of $N$ RIS elements, where $x_n$ denotes the binary phase shift of the $n$-th element, $x_n = 0$ and $x_n = 1$ represent a 0 and a $\pi$ phase shift, i.e., $\phi_n = 0$ and $\phi_n = \pi$, respectively.
Therefore the channel $\qh$ for the RIS-aided link between the BS and the UE can be formulated as follows:  
\vspace{-0.05cm}
\begin{equation}
\qh(\bm{x}) = \qh_\text{bu} + \qH_\text{br}\qPhi(\bm{x}) \qh_\text{ru}, 
\end{equation}
where correlated Ricean channels are considered for $\qh_\text{bu}$ and $\qh_\text{ru}$ as follows:
\vspace{-0.05cm}
\begin{equation}
\qh_{i} = \sqrt{\beta_i}\left( \sqrt{\frac{\kappa_i}{1+\kappa_i}} \qa_i +\sqrt{\frac{1}{1+\kappa_i}} \qR_i^{\frac{1}{2}} \qu_i \right), i \in \{\text{bu}, \text{ru}\},
\end{equation} 
where $\beta_i$ is the path gain, $\kappa_i$ is the Ricean $K$-factor, $\qa_i$ is the topology specific steering vectors, $\qu_i \sim \mathcal{CN} \left(\qzero, \qI\right)$, and $\qR_i$, $i \in \{\text{bu}, \text{ru} \}$, is the correlation matrix given in \cite{Singh2022}:  
\vspace{-0.1cm}
\begin{equation}
\begin{split}
& (\qR_{\text{bu}})_{vw} = \rho_\text{bu}^{d_{vw}/d_{\text{b}}}, \\
& (\qR_{\text{ru}})_{vw} = \sinc(2d_{vw}) \; \text{with} \; \sinc(2d_{\text{r}}) =  \rho_\text{ru}, 
\end{split}
\end{equation}
where $d_{vw}$ denotes the distance between the $v$-th and the $w$-th antenna/element at the BS/RIS. $d_{\text{b}}$ and $d_{\text{r}}$ denote the BS antenna and RIS element spacing, respectively. $\rho_\text{bu}$ and $\rho_\text{ru}$ denote the nearest neighbour BS antenna and RIS element correlations, respectively, and $0 \leq \rho_\text{bu} \leq 1$, $0 \leq |\rho_\text{ru}| \leq 1$.

A rank-1 LOS condition is considered between BS and RIS channel $\qH_\text{br}$ as follows:
\vspace{-0.05cm}
\begin{equation}
\qH_\text{br} = \sqrt{\beta_\text{br}} \qa_\text{b} \qa_\text{r}^H,
\end{equation}
where $\beta_\text{br}$ is the path gain between the BS and the RIS, $\qa_\text{b}$ and $\qa_\text{r}$ are the steering vectors at the BS and the RIS, respectively.

The received signal $\qy$ at the UE is given as follows:
\vspace{-0.05cm}
\begin{equation}
\qy(\bm{x}) = \qh(\bm{x}) s + \qn,
\end{equation}
where $s$ is the transmitted signal with power $Q$ and the noise vector $\qn \sim \mathcal{CN} \left(\qzero, \sigma^2 \qI \right)$.

Let $\qh(\bm{x}) = [h_1(\bm{x}), \dots, h_M(\bm{x})]^T$, where $h_m(\bm{x})$ denotes the channel between the $m$-th antenna at the BS and the UE with the RIS configuration $\bm{x}$, and $m = 1, \dots, M$. The signal-to-noise ratio (SNR) of the UE denoted by $\xi$ is written as
\vspace{-0.1cm}
\begin{equation}
\label{eqn:snr}
\xi(\bm{x}) = \frac{| \sum_{m=1}^{M} h_m (\bm{x})  |^2 Q} {\sigma^2}.
\end{equation}

The objective of the RIS optimization problem is to maximize the SNR of the UE by optimizing the phase shifts of the RIS elements. This problem involving $N$ integers is non-convex, hence challenging to obtain the solution, especially when N is large. It can be formulated as
\begin{equation}
\label{eqn:obj}
\begin{aligned}
\max_{\bm{x}} \; \; \xi(\bm{x}), 
\;\; \text{subject to} \;\;  x_{n} \in \{0,1\}, \; n = 1, \dots, N \,, 
\end{aligned}
\end{equation}

\vspace{-0.15cm}
Given that the received SNR is nearly proportional to $N^2$ when $N$ is large, and that the ratio between the variance and the mean squared SNR approximately follows the $1/N$ scaling \cite{Bjornson2021}, SNR optimization may also enhance channel hardening.

\vspace{-0.1cm}
\section{Cross-Entropy Algorithm for Phase Shifts Optimization in RIS-aided Communication}
\label{sec_ce}
\vspace{-0.1cm}
This work aims at optimizing a large number of discrete phase shifts in a moderate time complexity and evaluating its performance experimentally. The CE-based methods have been successfully applied to solve such discrete optimization problems in wireless communications, e.g., joint antenna selection \cite{zhang2010near}. Hence, the CE algorithm is considered to solve this RIS optimization challenge for the SNR improvement, which mainly depends on the Kullback-Leibler cross-entropy and the importance sampling in an iterative manner, each iteration includes four key phases: 1) generate a large number of random samples representing RIS configuration according to a specified probability distribution; 2) configure each sample to RIS and collect the corresponding performance, e.g., received SNR, in the absence of CSI; 3) sort the samples in descending order with respect to the performance indicator; and 4) update the distribution based on the top samples to generate better samples in the next iteration.
The detailed CE algorithm for the binary RIS optimization is given in Algorithm 1. 
To save time, if the CE algorithm has not yet converged to a binary vector at the end of the specified iterations, it will be enforced to be binary based on the probabilities as described in Step 5. 

Algorithm 1 presents a case of binary phase shifts due to the property of the RIS devices used for experiments. However, it can be simply extended to multi-phase scenarios. Let $C$ denote the number of phase shifts. For $C > 2$, the probability distribution $\textbf{P}^t$ will be a $C \times N$ matrix with $P_{cn}^t$ denoting the probability that the $n$-th RIS element to be the $c$-th phase at the $t$-th iteration. Since each element must be associated with only one of the phase shifts, the random samples may not meet this constraint where a projection strategy \cite{zhang2010near} is required to guarantee the feasibility of the random samples. Other steps for the multi-phase scenario will be the same as Algorithm 1.

\begin{algorithm}
\footnotesize
\caption{CE Algorithm for RIS Optimization.}  
\label{CE_algorithm}
\begin{algorithmic}
\STATE{\textbf{Step 1:}} Initialize the Bernoulli probability vector $\textbf{P}^0=\{ P_n^0 \}_{n = 1}^{N}$ with $P_n^0 = 0.5$, where $P_n^0$ denotes the initial probability of the $n$-th RIS element to be one. Set the total number of iterations as $T$ and the iteration index $t=1$.
\STATE{\textbf{Step 2:}} Randomly generate $K$ samples $\{\bm{x}^j\}_{j=1}^K$ according to Bernoulli distribution with probability $\textbf{P}^{(t-1)}$, where $\bm{x}^j = \{x_n^j\}_{n=1}^{N}$, and $x_n^j$ denotes the binary phase shift of the $n$-th element of the $j$-th sample.
\STATE{\textbf{Step 3:}} Commit each RIS configuration and collect the corresponding performance indicator, e.g., received SNR.

\Indp\FOR{$j =1:K$}
\Indp \item - Commit $\bm{x}^j$ as RIS configuration.
\Indp \item - Record the received SNR as $\xi_j$.
\ENDFOR

\Indm\STATE{\textbf{Step 4:}} Sort $\{\bm{x}^j\}_{j=1}^K$ in a descending order as $\{\bm{x}^{\sigma_j}\}_{\sigma_j=1}^K$ w.r.t. $\xi_j$. 

\STATE{\textbf{Step 5:}} Select the best $J = \lceil \varphi K \rceil$ samples from $\{\bm{x}^{\sigma_j}\}_{\sigma_j=1}^K$, where \\$\varphi$ denotes the quantile, e.g., $\varphi=0.1$, and then update the probability vector $\textbf{P}^{t}$ with
$P_{n}^{t} = \frac{\sum_{\sigma_j = 1}^{J} x_{n}^{\sigma_j} }{J}, \; n = 1, \dots, N \,.$

\Indp\IF{$\textbf{P}^t$ converges to a binary vector }
\Indp\STATE Break  
\ELSIF{$t = T$}
\Indp\STATE Post-process $\textbf{P}^t$ to be a binary vector with $P_n^{t} = 1$ \\if $P_n^{t} > 0.5$, otherwise $P_n^{t} = 0$, $n=1,\dots, N$.  
\ELSE
\Indp\STATE Set $t=t+1$ and go to Step 2.
\ENDIF

\Indm\STATE{\textbf{Step 6:}} Return $\textbf{P}^t$ as the optimized RIS configuration $\bm{x}^*$.

\end{algorithmic}
\end{algorithm}

Compared to the optimal solutions, the proposed CE algorithm reduces the computational complexity while achieving sub-optimal performance. With larger $T$ and $K$, the performance can be further enhanced but it will consume more time. Overall, it is a trade-off between performance and time complexity. However, this work mainly focuses on the experimental evaluation of the proposed algorithm. Therefore, an SNR improvement is favorable instead of the global optimum.


\vspace{-0.2cm}
\section{Simulation and Experiment Settings}
\label{sec_setting}

\vspace{-0.1cm}
\subsection{Simulation Settings}
\vspace{-0.1cm}
Note that the proposed CE algorithm for RIS optimization is model-free, which is not limited to any specific channel model. The channel model introduced in Section II and its parameters \cite{Singh2022} given in Table \ref{table:param_sim} are adopted only for simulation purpose.

\vspace{-0.21cm}
\begin{table}[ht]
\footnotesize
\caption{Main system parameters for simulations.}
\centering 
\begin{tabular}{|c |c|c |c|} 
\hline 
\textbf{Parameters} & \textbf{Values} & \textbf{Parameters} & \textbf{Values} \\ 
\hline
$\beta_\text{bu}$ & -81.7077 dB & $\kappa_\text{bu}$ / $\kappa_\text{ru}$ & 1 \\ 
\hline 
$\beta_\text{ru}$ & -67.0360 dB & $\rho_\text{bu}$ / $\rho_\text{ru}$ & 1 \\
\hline
$\beta_\text{br}$ & -34.1514 dB & $\sigma^2$ & -65 dBm \\
\hline 
\end{tabular}
\label{table:param_sim} 
\end{table}

The vertical uniform rectangular array in the y-z plane is considered for the steering vectors $\qa_i$ as follows \cite{Miller2019}:
\begin{equation}
\qa_i = \qa_{i,y} \otimes \qa_{i,z}, i \in \{\text{bu}, \text{ru}, \text{b}, \text{r}\}, 
\end{equation}
where $\otimes$ is the Kronecker product operation. Specifically, 
\begin{equation}
\begin{aligned}
&\qa_{i,y} = \left[ 1, e^{j2\pi \lambda_i \sin \theta_i \sin \omega_i}, \dots, e^{j2\pi \lambda_i (K_{i,y} - 1) \sin \theta_i \sin \omega_i  } \right], \\ 
&\qa_{i,z} = \left[ 1, e^{j2\pi \lambda_i \cos \theta_i}, \dots, e^{j2\pi \lambda_i (K_{i,y} - 1) \cos \theta_i } \right],
\end{aligned}
\end{equation}
and $K_{i,y}$ and $K_{i,z}$ denotes the number of antenna in the y and z axis, respectively, where  $K_{i, y}K_{i, z}= K_{i}$ and $K_i = M$ for $i \in \{\text{bu}, \text{b} \}$ and $K_{i}=N$ for $i=\{\text{ru, r}\}$. $\theta_i$ and $\omega_i$ are elevation/azimuth angles of arrival or angles of departure. $\lambda_i$ is the wavelength, where $\lambda_\text{bu} = \lambda_\text{b} = 0.5$ and $\lambda_\text{ru} = \lambda_\text{r}$ satisfying $\text{sinc} (2 \lambda_\text{r}) = d_\text{r}$, with the nearest neighbour RIS element correlation $0\leq |d_\text{r}| \leq 1$. The following parameters are used throughout the simulations: $\theta_\text{b} = 109.9^{\circ}$,  $\omega_\text{b}= -29.9^{\circ}$, $\theta_\text{r} = 77.1^{\circ}$, $\omega_\text{r}= 19.95^{\circ}$, $\theta_\text{bu} = 80.94^{\circ}$, $\omega_\text{bu}= -64.35^{\circ}$,  $\theta_\text{ru} = 71.95^{\circ}$, $\omega_\text{ru}= 25.1^{\circ}$  \cite{Singh2022}. 

\vspace{-0.3cm}
\subsection{Experiment Setup}
\vspace{-0.1cm}
All experiments are conducted in the static RC which provides a multipath fading environment as shown in Fig. \ref{sys_model}. The benefit of testing the RISs in RC, not in free-space, is that the backscattering from the cavities changes the EM reflective response of the RISs themselves, hence the model-free optimization works as if the RISs operate in arbitrary indoor environments.
The BS and the UE are modeled and implemented by the Universal Software Radio Peripheral (USRP) X310 devices. 
Main parameters used for USRP are listed in Table \ref{table:param_usrp}. RIS devices from Greenerwave are used in the experiments, and each RIS device has $N_0 = 76$ configurable elements. Detailed introduction about RIS is given in \cite{gros2022multi}. 
The proposed algorithm is executed on a laptop at the receiver side which connects to RIS devices via USB cables to control them.

\vspace{-0.2cm}
\begin{table}[ht]
\footnotesize
\caption{Main parameters of the transceiver (USRP).}
\centering 
\begin{tabular}{|c |c|} 
\hline 
\textbf{Parameters} & \textbf{Values} \\ 
\hline
Center Frequency & 5 GHz  \\ 
\hline 
Master Clock Rate & 200 MHz \\
\hline
Interpolation / Decimation Factor & 100 \\
\hline 
Tx Gain & 10 dB \\
\hline 
Rx Gain & 10 dB \\
\hline 
\end{tabular}
\label{table:param_usrp} 
\end{table}

\vspace{-0.7cm}
\section{Numerical and Experimental Results}
\label{sec_result}
\vspace{-0.3cm}
This section provides both numerical and experimental evaluations for the proposed CE algorithm in RIS-aided wireless networks. Unless otherwise stated, for both simulations and experiments, $M = 2$, $N = 152$, $T = 15$, $K = 100$, $\varphi = 0.1$ and the transmit gain of 10 dB are adopted for performance evaluations, and the average SNR is given by 1000 collected SNR values in the following results.
The proposed CE algorithm is compared with the following benchmarks:
\begin{itemize} 
\item \textbf{SA} \cite{mohammed2022reconfigurable}: This method mainly relies on the MC sampling, where the initial temperature and the final temperature are set to be 100 and $10^{-8}$, respectively, and the cooling rate is set to be 0.995. 
\item \textbf{MH} \cite{deng2021metropolis}: In addition to the MC random sampling, this method adopts an acceptance ratio between the SNRs achieved by the new sample and the current sample to decide whether to accept or reject the new candidate.
\item \textbf{Conditional Sample Mean (CSM)} \cite{ren2022configuring}: This method depends on the average performance of a large number of random samples conditioned on each phase shift of each RIS element.
\item \textbf{Without RIS}: This scenario acts as a lower bound.
\end{itemize}

For fair comparisons, the CE algorithm that has $T$ iterations with $K$ random samples per iteration is equivalent to the benchmarks with $TK$ samples, and they are uniformly called $T$ iterations in the following sections.

Note that the simulations are performed based on the Ricean channel model introduced in Section IV-A. For experiments, the channel in the RC is a general Ricean channel emulating the scenario in simulations.

\vspace{-0.2cm}
\subsection{Case Study on Iterations}
\begin{figure}[!htb] 
\centering  
  \begin{subfigure}{0.38\textwidth}
    \centering\includegraphics[trim=2.5cm 0cm 2.5cm 1cm,clip,width=\linewidth]{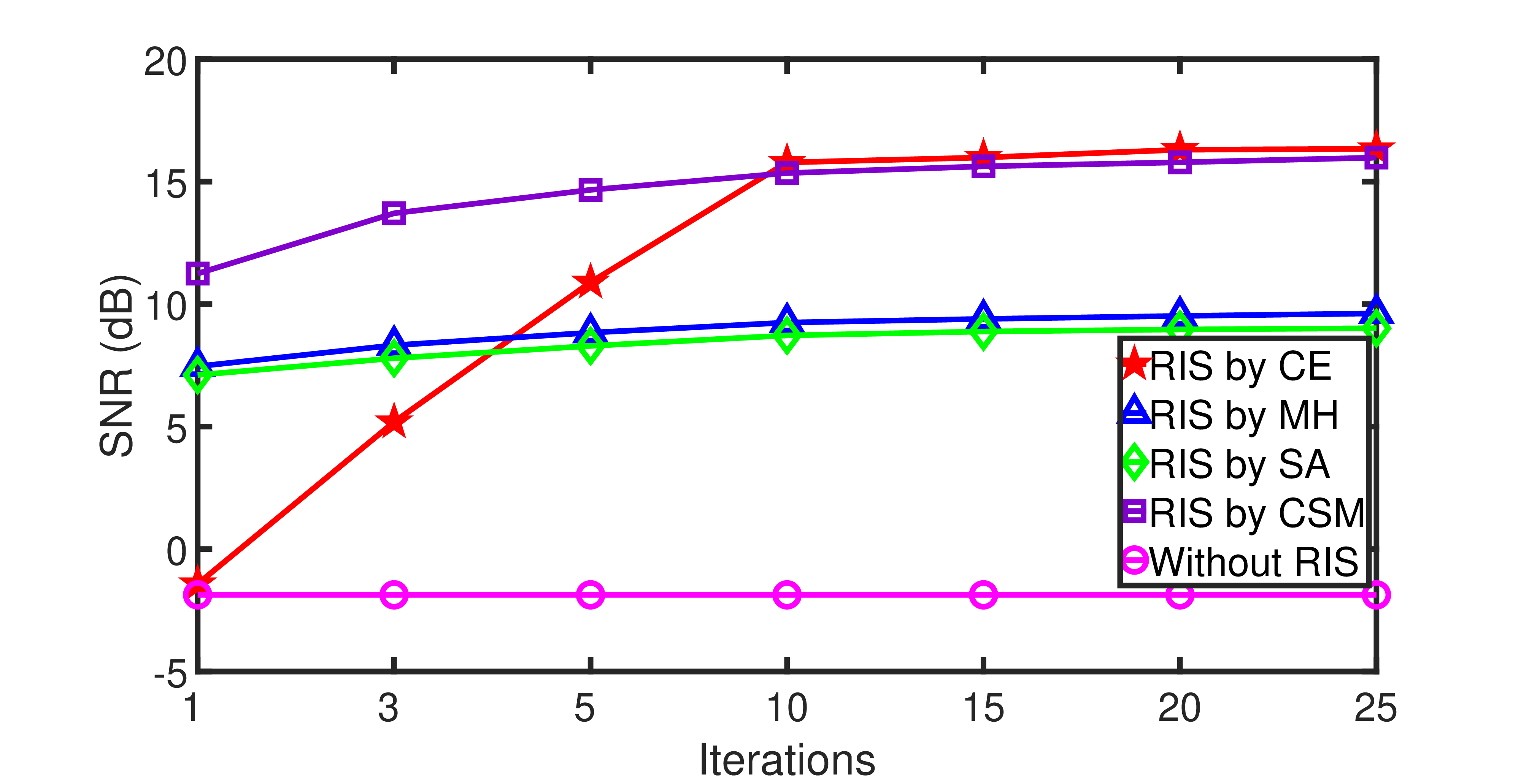} 
    \caption{Simulations.}
    \label{sim_ite}
  \end{subfigure}
  \begin{subfigure}{0.38\textwidth}
    \centering\includegraphics[trim=2.5cm 0cm 2.5cm 0cm,clip,width=\linewidth]{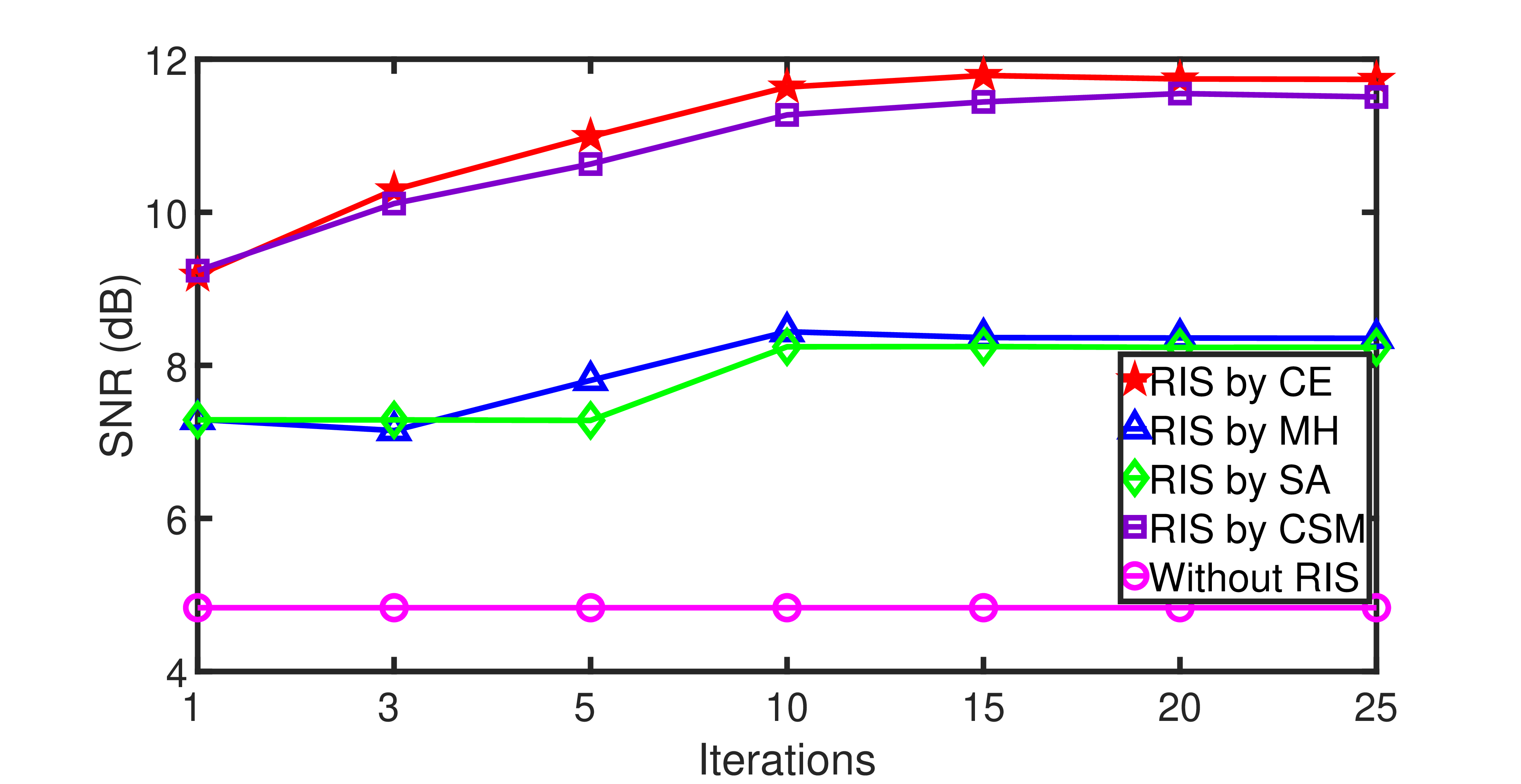}
    \caption{Experiments.}
    \label{exp_ite}
  \end{subfigure}
\caption{Performance on different numbers of iterations.}
\label{result_ite}
\end{figure}

\vspace{-0.1cm}
The performance of the proposed CE algorithm and the benchmarks is studied on different numbers of iterations as illustrated in Fig. \ref{result_ite}. 

\vspace{-0.45cm}

\vspace{-0.2cm}
For the simulations shown in Fig. \ref{sim_ite}, the average SNR is only -1.9 dB without RIS. The proposed CE algorithm increases the SNR to around 16 dB, which outperforms the MH, the SA and the CSM benchmarks when the CE converges (from 10 iterations).
From experiment results shown in Fig. \ref{exp_ite}, the average SNR is only 4.8 dB without RIS, while the CE algorithm enhances the SNR to nearly 12 dB, which indicates superior SNR improvement than the benchmarks.
The difference in channel models between the simulations and the experiments in the RC will result in performance differences in the early iterations, which can be observed in Fig. \ref{result_ite}.
Both simulation and experiment outcomes demonstrate that the proposed CE algorithm brings a better quality of service for the end-user.

\vspace{-0.28cm}
\subsection{Channel Hardening}
\vspace{-0.1cm}
Channel hardening involves changing the number of RIS elements. However, practical RIS devices do not allow to configure only parts of elements due to hardware limitations. In the experiment, $N^* = \lceil N / N_0 \rceil$ RIS devices are used, and if $N < N_0 N^*$, then $N_0 N^* - N$ RIS elements will be set to be zeros in order to imitate that $N$ elements are optimized.

\begin{figure}[!htb] 
\centering
  \begin{subfigure}{0.38\textwidth}
    \centering\includegraphics[trim=2.5cm 0cm 2.5cm 1.2cm,clip,width=\linewidth]{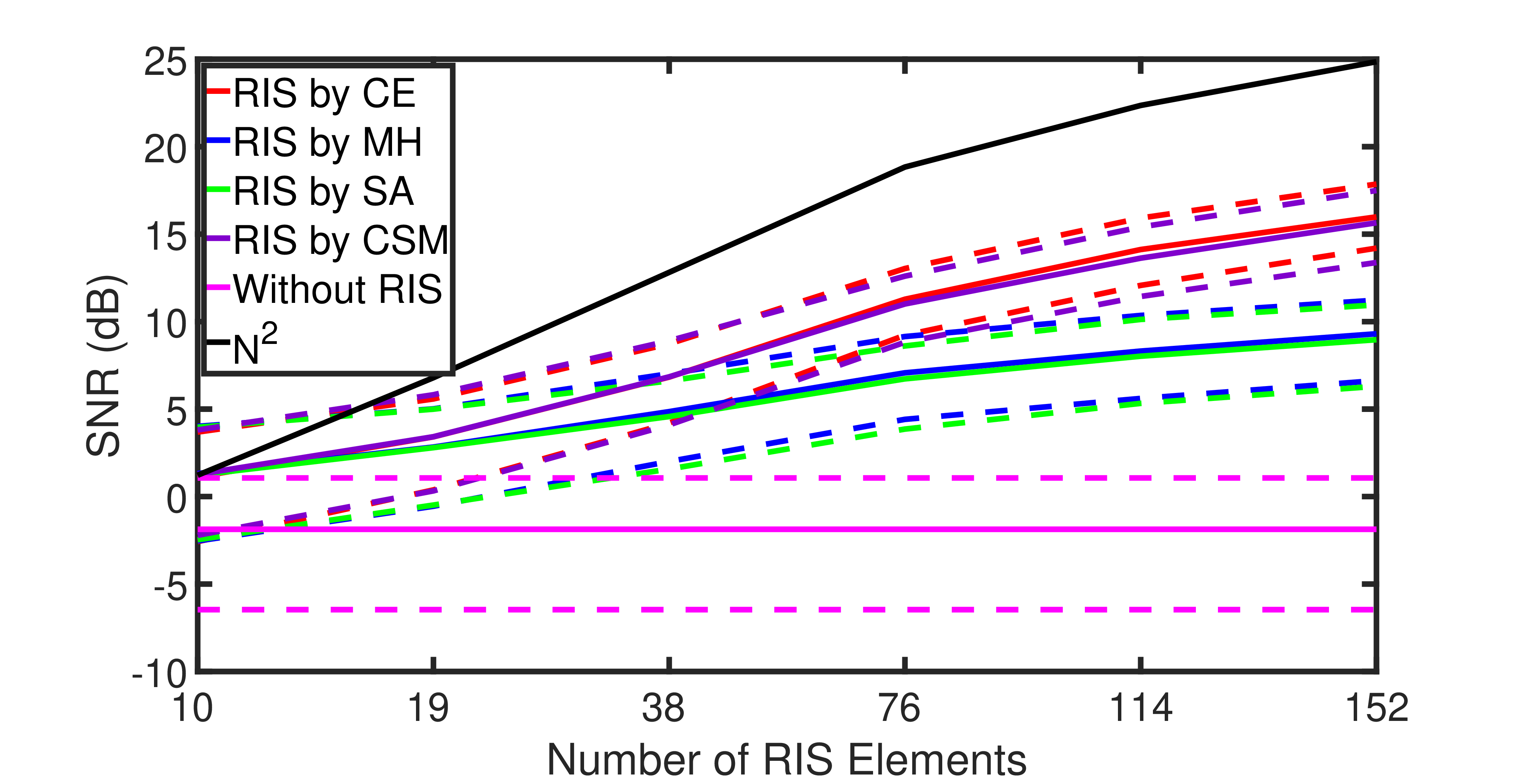} 
    \caption{Simulations.}
    \label{sim_ch}
  \end{subfigure}
  \begin{subfigure}{0.38\textwidth}
    \centering\includegraphics[trim=2.5cm 0cm 2.5cm 0cm,clip,width=\linewidth]{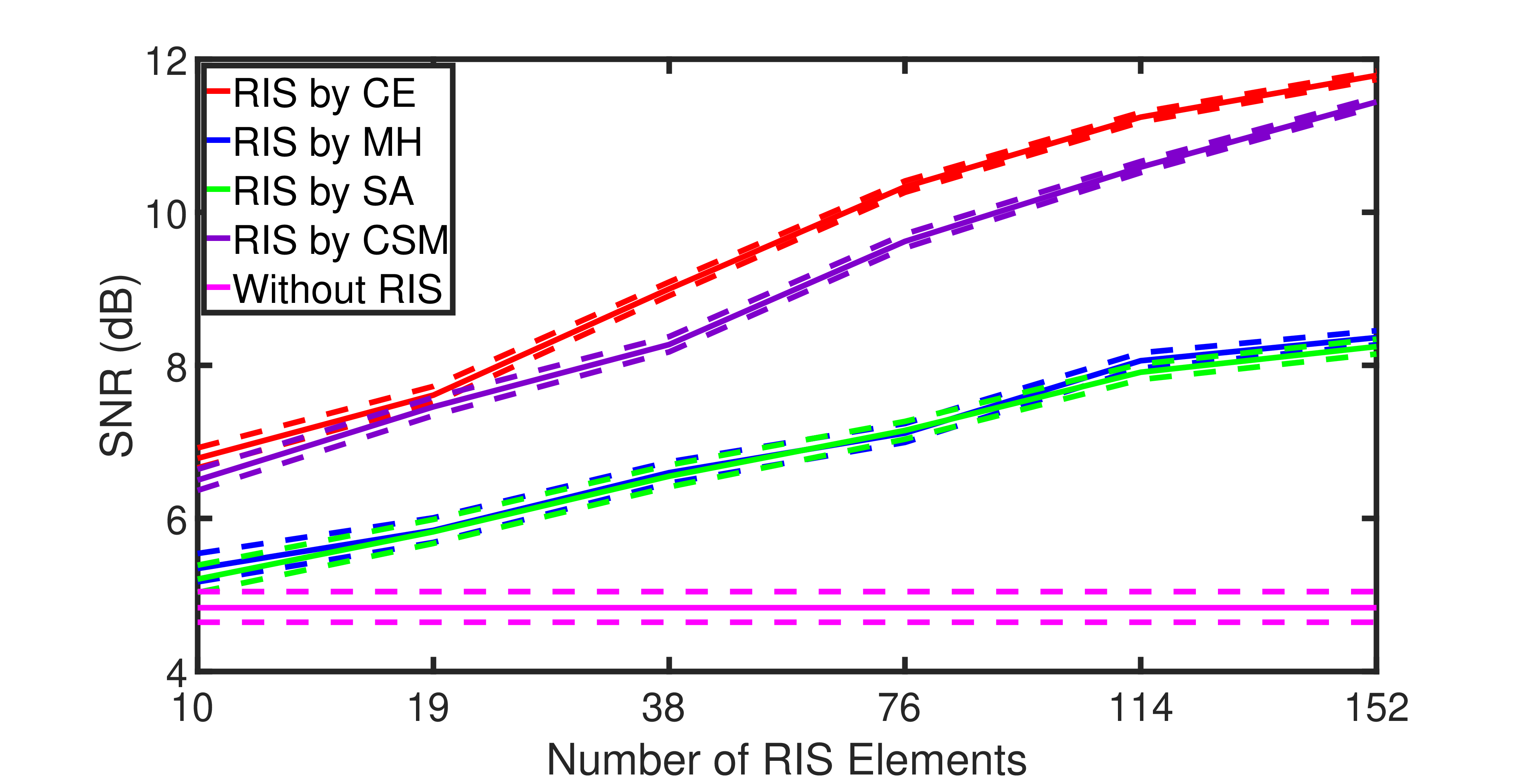}
    \caption{Experiments.}
    \label{exp_ch}
  \end{subfigure}
\caption{Performance on channel hardening.}
\label{result_ch}
\end{figure}

The channel hardening performance of the proposed CE algorithm is compared against the benchmarks as shown in Fig. \ref{result_ch}, where the results are given by the average SNR (solid lines), 10\% and 90\% quantiles of 1000 collected SNRs (dashed lines) for measuring the channel hardening effects.

Fig. \ref{result_ch} validates that the average SNRs of all algorithms improve with an increasing number of RIS elements. The proposed CE method outperforms the benchmarks in terms of both SNR improvement and channel hardening, and it shows greater advantages than the benchmarks with growing numbers of RIS elements. Additionally, the practical experiments achieve stronger channel hardening than the simulations.

In Fig. \ref{sim_ch}, a curve of $N^2$ is also given, which demonstrates that the CE algorithm achieves a similar trend as indicated by the square law expectation \cite{Bjornson2021} between the SNR and the $N^2$. Despite the fact that the SNR improvements in the experiments in Fig. \ref{exp_ch} slightly deviate from the square law due to the potential influence of imperfect RIS element deactivation, the CE approach leads to significant SNR enhancement and channel hardening.

\begin{figure}[!htb] 
\centering
  \begin{subfigure}{0.38\textwidth}
    \centering\includegraphics[trim=2cm 0cm 2.5cm 0cm,clip,width=\linewidth]{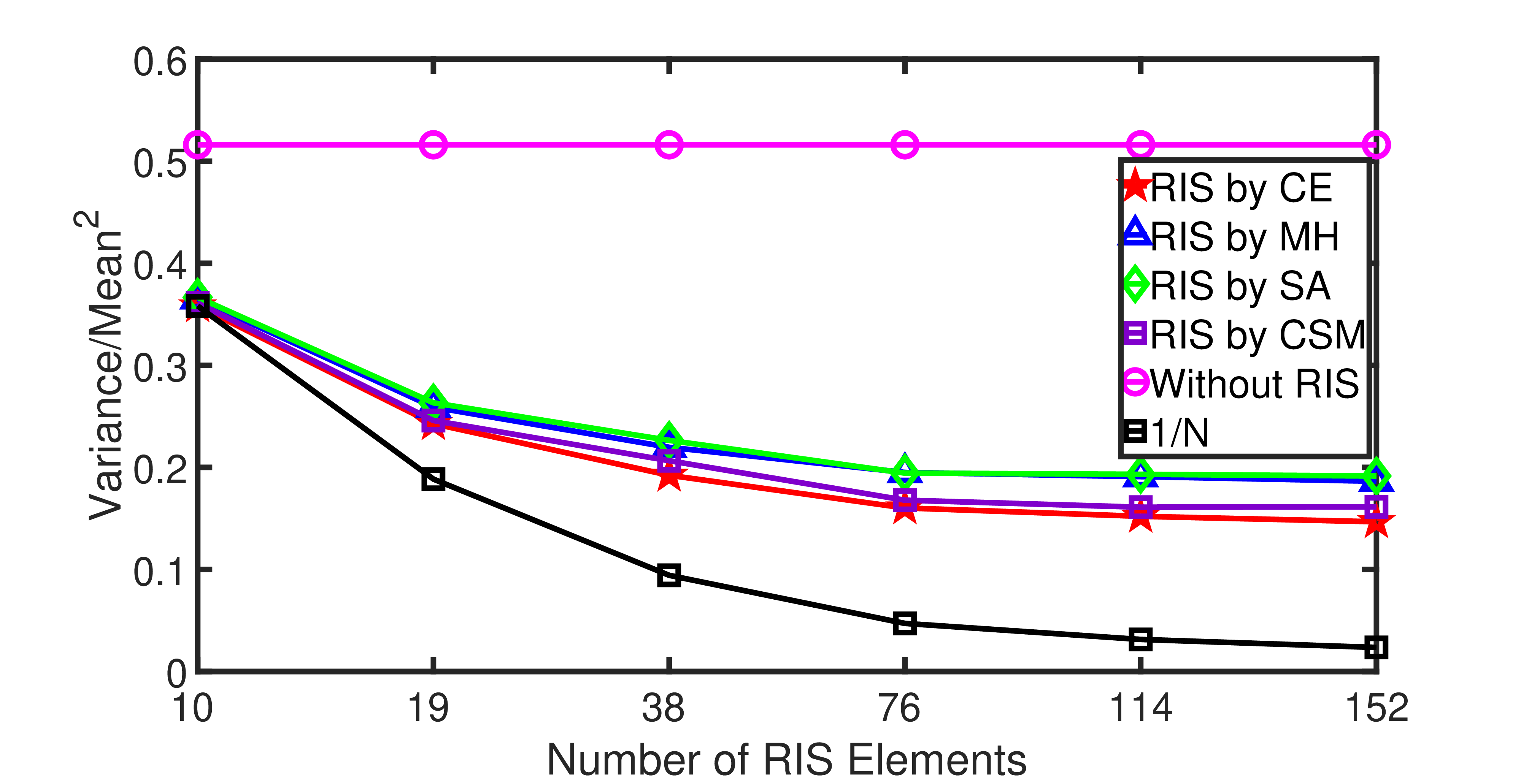} 
    \caption{Simulations.}
    \label{sim_ch_ratio}
  \end{subfigure}
  \begin{subfigure}{0.38\textwidth}
    \centering\includegraphics[trim=2cm 0cm 2.5cm 0cm,clip,width=\linewidth]{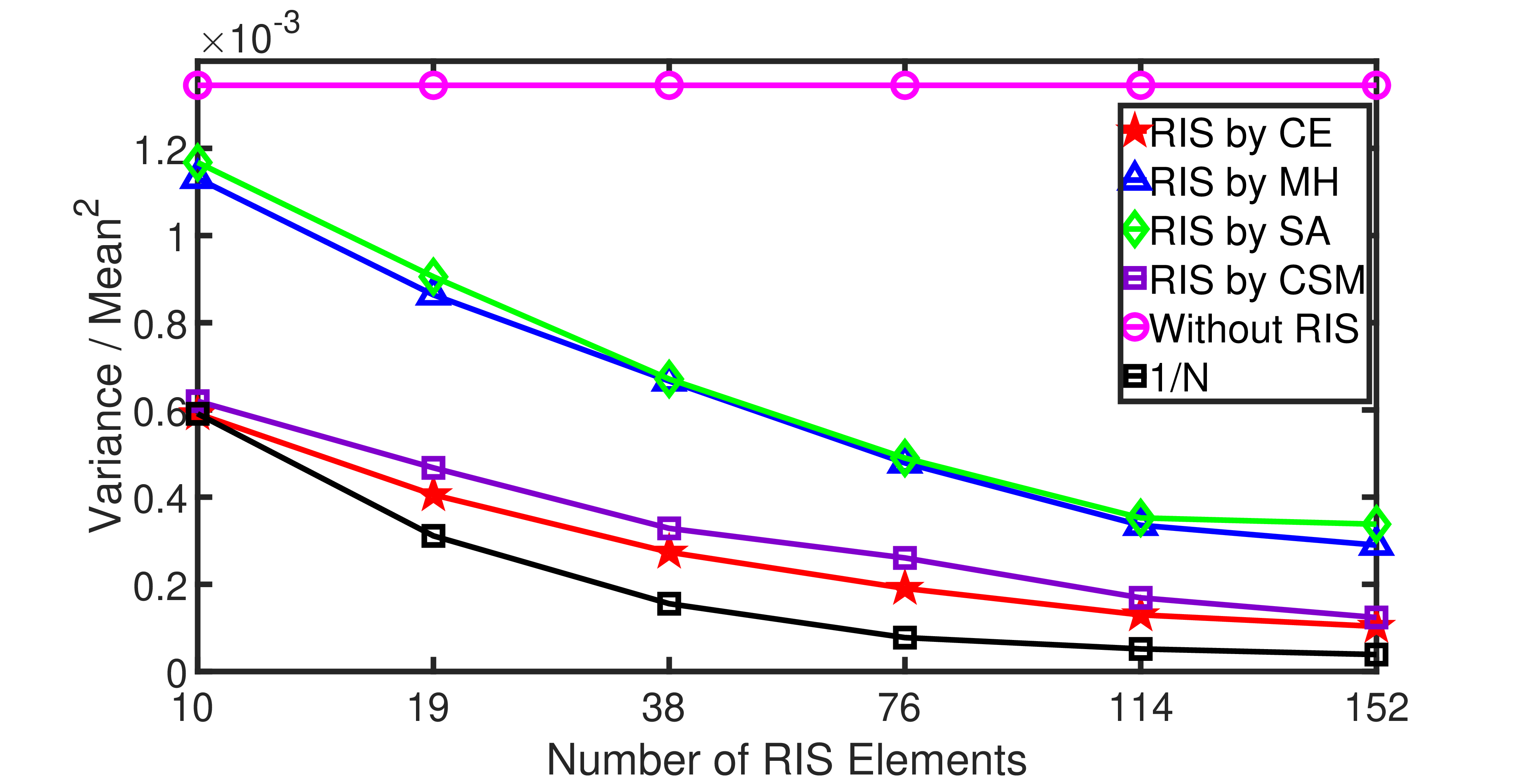}
    \caption{Experiments.}
    \label{exp_ch_ratio}
  \end{subfigure}
\caption{Performance on channel hardening (variance$/$mean$^2$).} 
\label{result_ch_ratio}
\end{figure}

In order to further evaluate the channel hardening performance, Fig. \ref{result_ch_ratio} demonstrates the ratio between the variance and the mean square of the collected SNRs with increasing numbers of RIS elements. Both simulation and experiment results in Fig. \ref{result_ch_ratio} indicate that the ratio generated by the CE algorithm and benchmarks generally keeps decreasing with larger $N$, and the proposed CE algorithm remains the lowest, e.g., for $N = 152$, the ratio is approximately 0.15 and $ 1 \times 10^{-4}$ for simulation and for experiment, respectively, which illustrates that the channel hardening performance of the CE algorithm is superior to the benchmarks. Additionally, the performance of the CE algorithm generally follows the trend of the curve of the theoretical scaled $1/N$.

\vspace{-0.1cm}

\vspace{-0.2cm}
\subsection{Time Complexity}
\vspace{-0.1cm}
The execution time of the proposed CE and the benchmarks is compared for $N=152$, $T=15$ and $K=100$ as shown in Table \ref{table:time}, where all algorithms show a similar level of time performance for the simulation and the experiment, respectively. The reason is that they share the same complexity of approximately $ \mathcal{O}(TKN) $. For the simulation, the execution time of the proposed CE algorithm is 49 ms, which is close to the MH and the SA, and achieves nearly 20 ms faster than the CSM.
For the experiment, the execution time is given by the overall time induced by the optimization algorithms, which are ranging from 46 s to 48 s approximately, wherein configuring random samples to RIS devices consumes in total 21 s, which is also decided by hardware.
Although the CSM achieves comparable SNR boost and channel hardening as the CE algorithm, the proposed CE algorithm requires less execution time.

\vspace{-0.15cm}
\begin{table}[!htb]
\centering
\small
\caption{Overall execution time for $N=152$, $T=15$ and $K=100$ in seconds.}
\begin{tabular}{|c|c|c|c|c|c|}
\hline
Algorithms & CE & SA & MH & CSM  \\ \hline
Simulation & 0.0490 & 0.0453 & 0.0449 & 0.0683 \\ \hline
Experiment & 46.0229 & 45.6679 & 45.6322 & 48.3675 \\ \hline
\end{tabular}
\label{table:time}
\end{table}

\vspace{-0.6cm}
\section{Conclusion}
\label{sec_con}
\vspace{-0.1cm}
This work proposes a CE optimization algorithm for the RIS-aided communication system. It only adopts received SNR as performance feedback and does not require CSI. Besides, it is model-free and applicable to scenarios of arbitrary channel models.
Proved by simulations and experiments, the proposed CE-based RIS optimization algorithm outperforms the benchmarks with better end-user SNRs, while maintaining similar time complexity. Additionally, the proposed CE algorithm achieves stronger channel hardening effects than the benchmarks, which is promising for channel stability.
Moreover, the proposed CE algorithm can be easily extended to multi-phase RIS optimizations for broad applications. It can be implemented in the multiuser scenario and will be studied in the future work.

\vspace{-0.1cm}
\section{Acknowledgement}
\vspace{-0.1cm}
We are grateful to Anas Al Rawi (OFCOM) for the insights on the problem of RIS-assisted Channel Hardening.

\bibliographystyle{IEEEtran}
\bibliography{ZHANG_WCL_ref}
\end{document}